\documentclass[twoside]{article}
\usepackage{fleqn,espcrc2,epsf}

\newcommand{\AmS}{{\protect\the\textfont2
  A\kern-.1667em\lower.5ex\hbox{M}\kern-.125emS}}

\newif\ifcolor
\colortrue

\title{Comparing Clusters and Supercomputers for Lattice QCD
}

\author{
        Steven Gottlieb\\
{Department of Physics, Indiana University, Bloomington, IN 47405, USA}
} 
       
\begin{document}

\begin{abstract}
Since the development of the Beowulf project to build a parallel
computer from commodity PC components, there have been many such
clusters built.  The MILC QCD code has been run on a variety of
clusters and supercomputers.  Key design features are identified,
and the cost effectiveness of clusters and supercomputers are
compared.  

\end{abstract}

\maketitle

\section{INTRODUCTION}
The Beowulf project began in 1994 at the NASA Goddard Space Flight Center.
More about the history of the project and the current status can be found
at the web site http://www.beowulf.org.  There are about 100 clusters
listed on the Beowulf home page (which is clearly not a complete list).
Within the MILC collaboration, we have access to at least five
clusters at our universities.  We also have done production work on six
other clusters  
at national supercomputer centers.

There are several advantageous characteristics often cited for clusters.
Chief among these is the use of commodity hardware to produce a very
cost-effective computer.  Processors such as the Intel Pentia and Celeron
and the AMD Athlon or K6, FastEthernet network cards and switches
have been used to build quite cost-effective machines.  However, other
choices such as the Compaq Alpha processor and higher speed networks
such as Myrinet, Giganet and Quadrics QsNet have also been used to build very
powerful clusters.  Another characteristic of clusters is the use of
commodity software such as Linux, GNU, MPICH and PBS to keep software
costs close to zero.  A third advantage of the cluster approach is their
programmability and flexibility.  Message Passing Interface or MPI, has
become a standard in commercial parallel computers.  The MILC
code had been compiled under MPI well before being run on a Beowulf
cluster.  The port to Beowulf required minimal effort.  All of our
PC cluster benchmarks have been done without any assembly code.  Practical
calculations can be done on current clusters with a granularity that
is well suited to FFT routines.  Clusters have a community of
users and developers.  New system administration tools frequently become
available, as do advances in parallel file systems, schedulers and other
useful software.  Thus, one can take advantage of the vigor of the
community and avoid spending a large amount of time developing software
unrelated to the physics.  Finally, because of the short design time
of clusters, one can take quick advantage of the many developments in PC
hardware.  It is not necessary to lock oneself into a technology either
well in advance of it's actually being available, or a well developed
technology that will be outdated by the time a large system can be
constructed and commissioned.  One can often avoid the problem of having
a single source for key items.  If you can no longer get a particular
motherboard, there
will be another vendor with a similar (or superior) offering.

There are also potential disadvantages of clusters.  With a standard
supercomputer, one can get a maintenance contract, and there is somebody
to yell at when things go wrong.  (However, as a long-time user of
supercomputers, I know that having a vendor doesn't assure that
the problem will be fixed.)  Recently, a number of vendors have been
selling clusters.  So the problem of not having anyone to yell at may be
avoided.  Of course, there is still no assurance that yelling (or even asking
politely) will result in the problem being solved.
Another disadvantage of the cluster approach is that of having to rely on
the design effort of others.  If vendors are not producing hardware with
the specifications that you need, you may not be able to build a well
optimized system.  On the other hand, most
physicists are neither skilled at nor interested in VLSI design or PCB
layout and would rather spend their time thinking about physics, so why not
take advantage of the labor of computer engineers?


The Indiana University Physics Department received \$50,000 in 1998
to build a 32-node Linux cluster.  The machine we built is called
CANDYCANE, which stands for CPUs And Network Do Your Calculation And
Nothing Else.  CANDYCANE is an appropriate name because it was designed for
the ``sweet spot,'' that is, components were picked to give the best 
price-performance ratio attainable.  It is used by several research groups in the
department, but usually only one or two jobs are running at the same time.
In September 1998, a four-node prototype cluster was built and tested.  
Three different ethernet cards were tested to see if the higher priced
cards could be justified by superior performance.  Detecting no difference
in performance, we selected the least expensive card for the production
cluster.  In October, the purchasing department put out a request for bids 
on the desired components.  In November, just before Thanksgiving, the last
of the components arrived.  (Several vendors were used to get the best
price on each component.)  On the Wednesday and Friday of Thanksgiving
break, 34 nodes were built, the software was installed and everything was
placed on shelves and connected.  One node serves as a console, and one
as a spare.  The cost per node was \$693 for a Pentium II 350, with a 4.3
GB hard drive and 64 MB of ECC RAM.  Each node has a floppy drive and a
FastEthernet card.  However, the compute nodes have no keyboard, video card
or CDROM.  There are a few video cards and an extra keyboard that can be used
if a node does not reboot on its own.  The 40-port HP Procurve switch cost
about \$2,000, so the total cost was about \$25,000.  Currently, 
(October, 2000)
it would be possible to build this system for $<$\$300 per node, or
for approximately \$11,500.  An
even more attractive alternative would be a diskless Athlon 600 MHz
system for which the per node cost is about \$250.  This node would have
much better performance than the PII 350; however, the FastEthernet would
be a bottleneck on the MILC code with Kogut-Susskind quarks.  Even so,
a 32 node system with a minimum performance of 1280 and 1660 Mflops, for $8^4$
and $14^4$ sites per node, respectively, could be built for under \$10,000.
This works out to a cost/MF of between \$6.00 and \$7.80.

In Sec.~2, we describe the key issues for good performance and in
Sec.~3, we present benchmarks for the MILC code on various supercomputers
and clusters.  Section 4 gives rough cost-performance ratios for a number
of platforms.  
Additional information about emerging technologies for clusters and user
experiences can be found in the on-line presentations from a session that
I organized at the March, 2000 APS meeting \cite{APS}.

\section{KEYS TO PERFORMANCE}

A very simple approach to achieving good performance for domain
decomposition codes like Lattice QCD codes is to optimize single node
performance and to try to avoid degrading performance too much when one has to
communicate boundary values to neighboring nodes.

The single node performance is likely to depend upon such issues as the
quality of the CPU, the performance and size of cache(s), the bandwidth to
main memory and the quality of the compiler.  For message passing
performance, key issues are the latency,
peak bandwidth, processor overhead and the message passing software.

\begin{figure}[t]
\epsfxsize=0.99 \hsize
\epsffile{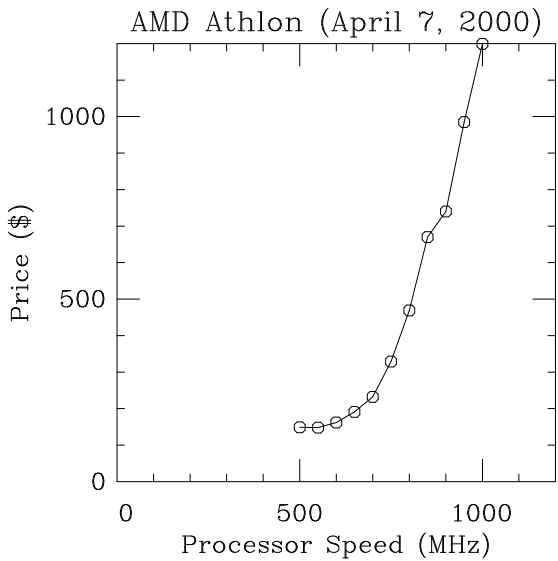}
\vspace{-28pt}
\caption{Processor price {\it vs}. speed.}
\label{fig:price_vs_speed}
\end{figure}

\begin{figure}[thb]
\epsfxsize=0.99 \hsize
\epsffile{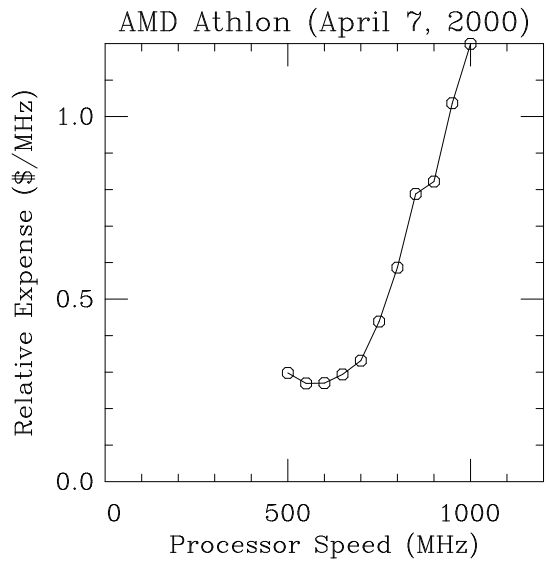}
\vspace{-28pt}
\caption{Dollars per MHz {\it vs}. speed.}
\label{fig:value}
\end{figure}

Focusing first on single node performance, we note that it is easy to waste
a lot of money on a poor system design.  To illustrate this, we consider
the various speed AMD Athlon processors available and their prices on a
particular day.  Although we focus on Athlon here, the same considerations
apply to Intel or other processors.  Figure \ref{fig:price_vs_speed}
shows that processor price
is a rapidly increasing function of speed.  In Fig~\ref{fig:value}, 
we divide the price
by the speed of the chip and see that the relative expense rises
rapidly for the faster chips.  At the time this graph was produced, there
was an apparent sweet spot at 600 MHz.  The faster chips have a higher
price-performance
ratio.  Depending upon the costs of the other components of the system,
the entire system may have a higher or lower price-performance ratio.

For our QCD codes, access to memory is quite important.  With the benchmarks 
below we demonstrate that performance does not increase in proportion to the
speed of the chip.  This is because memory speed is fixed by the 100 MHz
Front Side Bus for both 500 MHz and 600 MHz Athlons.

\begin{table}[bht]
\caption{Megaflop rate of Athlon Processors}
\label{tab:athlon}
\begin{tabular}{ccc}
\hline
L & 500 MHz & 600 MHz \\
\hline
4 & 231 & 276 \\
6 & 129 & 135 \\
8 & 97 & 102 \\
10 & 92 & 97 \\
12 & 90 & 95 \\
14 & 89 & 93 \\
\end{tabular}
\end{table}
\begin{table*}[thb]
\caption{Megaflop rates of various motherboards and CPU combinations or cluster nodes}
\label{tab:pc133}
\begin{tabular}{ccccccc}
\hline
L & Gigabyte GA6VXE+ & Intel CC820 & SM PIIISED & SM P6SBA 
& RR \dag & LL \ddag \\
  & Pentium III 533B    & PIII 533B & PIII 533B & PII 350 
& PII 450 & PIII 733 \\
\hline
4 & 186 & 182 & 174 & 114 & 142 & 319 \\
6 & 106 & 98 & 94 & 83 & 99 & 140 \\
8 & 81 & 75 & 73 & 72 & 82 & 130 \\
10 & 76 & 72 & 70 & 70 & 79 & 127 \\
12 & 76 & 70 & 69 & 70 & 78 & 127 \\
14 & 73 & 70 & 69 & 70 & 78 & 126 \\
\hline
\multicolumn{3}{l}  \dag Roadrunner: Portland Group compiler \hfil\\
\multicolumn{3}{l} \ddag Los Lobos: Portland Group compiler \hfil\\
\end{tabular}
\end{table*}

The 600 MHz chip has a peak speed 20\% faster than the 500 MHz chip.  With
$4^4$ lattice points, we do see a 20\% speed up, but for the larger problem
sizes that do not fit into cache, there is only a 5\% speedup.  We expect
that for even faster processors performance increases will be marginal.

Since memory access is so crucial, I have purchased a Pentium III 533B
chip that uses PC133 memory.  In theory, it should provide about 33\%
better performance than a similar chip with PC100 memory.
I have tried three different
motherboards using different support chips and the results are
disappointing.
The Gigabyte GA6VXE+ motherboard uses a VIA chip set, 
the Supermicro (SM) PIIISED
uses the Intel 810e chip set and I also tried an Intel CC820 motherboard
using the Intel 820 chip set.  The results are not significantly better 
than a PII
350 chip using a BX motherboard  and the GNU C compiler
and are worse than a PII 450 using the Portland Group C compiler.
(See Table~\ref{tab:pc133}.)

However, there are support chips from ServerWorks that support PC133
memory well.  Here are results from Los Lobos (LL), that uses 
Intel 733 MHz chips in IBM Netfinity servers that use the ServerWorks chip set. 
SuperMicro is manufacturing a dual processor
motherboard that uses a support chip from ServerWorks, but the motherboard
currently costs about \$280, which is about twice the price of
dual processor BX motherboard.  Also, this board requires registered memory
which will add to the cost of the system.

One of the disadvantages of using Athlon rather than Pentium chips is that
there are not dual or quad processor boards available for the
Athlon.  However, this should change soon as in early October, 2000 AMD
demonstrated a dual processor system using double data rate (DDR) memory.

Turning now to multiprocessor performance, we find that a simple
performance model of the Kogut-Susskind Conjugate Gradient
algorithm gives this bandwidth requirement to overlap communication and
floating point operations:

\begin{equation}  
MB = 48 MF/ (132 L) = 0.364 MF / L ,
\end{equation}

where $MB$ is the {\it achieved\/} bandwidth in Megabyte/s, $MF$ is the 
{\it achieved\/} floating
point speed in Megaflop/s on matrix-vector multiplication
and an $L^4$ portion of the grid is on
each node.  We assume there are neighboring nodes in each direction, {\it
i.e.}, 16 or more nodes.  The constant factor 0.364 is specific to
KS quarks.  However, the $1/L$ behavior is typical of the domain decomposition
approach to parallelism and comes from the surface to volume ratio.

\begin{figure}[t]
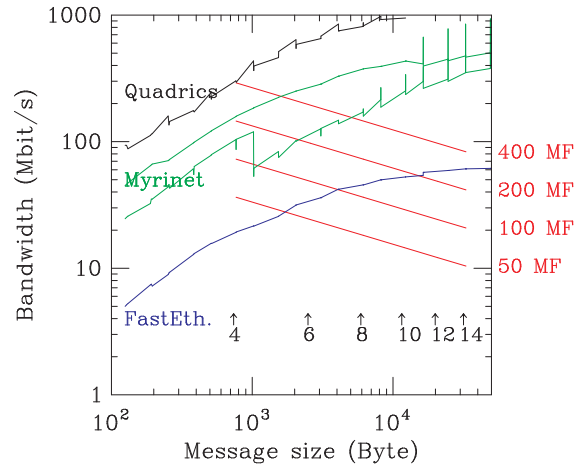

\ifcolor{\epsfxsize=0.99 \hsize\epsffile{fig3.ps}}
\else{\epsfxsize=0.99 \hsize\epsffile{fig3bw.ps}}\fi
\vspace{-28pt}
\caption{Measured bandwidth and the simple performance model.}
\label{fig:performance}
\end{figure}

Figure \ref{fig:performance} shows a log-log plot 
of measured bandwidth on a ping-pong test for
three types of hardware and the performance model for several processor speeds.
The messages vary in size from 800 bytes to 30 KB for problem sizes of
interest.  The arrows near the bottom of the graph correspond to
different L values.
The FastEthernet \ifcolor{(blue)\ }\fi and Myrinet \ifcolor{(green)\ }\fi
curves come from measured performance on the Roadrunner (RR)
supercluster at the
Albuquerque High Performance Computer Center. 
Two curves are shown for Myrinet.  With the newer drivers, bandwidth is better
and smoother.
The Quadrics curve comes from the Teracluster at Lawrence Livermore
National Laboratory (LLNL).
The measurement was done using the Netpipe program from the
Ames Scalable Computing Laboratory \cite{AMESLAB}.  
The straight \ifcolor{red\ }\fi lines come from the 
performance model presented above and
are plotted for matrix times vector speeds of 50, 100, 200 and 400 MF.
We need to run at a large enough value of L so that the measured bandwidth
is above the red line (for what ever speed our processor achieves for
the corresponding value of L).  Because of cache effects, the processors
will achieve higher speeds when $L$ is small, but that requires the highest
bandwidth.  Thus, pushing up the communication rate for small
messages is important.  Being able to run for a small value of $L$ with
high efficiency allows running a fixed size problem at high total
performance.
We see that none of the networks achieves more than a small fraction of its peak
bandwidth for the message sizes of interest.  A system design based on
achieving that peak bandwidth would almost certainly be communication
bound, {\it i.e.}, money would have been spent on floating point capacity
that could not be used.  There are large differences in the prices of
FastEthernet, Myrinet and Quadrics hardware.  Choice of network can
obviously play a critical role in system performance and
cost-effectiveness.

In addition to producing the data for the bandwidth curves, Netpipe provides 
the latency.  It is easier to remember the
latency of the different hardware and software combinations than the full
bandwidth curve, so we tabulate that here.

\begin{table}[bt]
\caption{Latency}
\label{tab:latency}
\begin{tabular}{lc}
\hline
Hardware/software & Latency ($\mu$s)\\
\hline
CANDYCANE MPICH & 151-166 \\
Roadrunner MPICH & 169-179 \\
CANDYCANE MVICH & 60 \\
CANDYCANE GAMMA & $\approx$ 41\\
\hline
Roadrunner Myrinet & 31-34 \\
Roadrunner Myrinet${}^*$ & 16 \\
\hline
Teracluster Quadrics & 7\\
\hline
${}^*$latest drivers &\\
\end{tabular}
\end{table}

MPICH with FastEthernet clearly has the longest latency.  Unfortunately, the
larger value is for a more recent kernel.  The longer latency for Roadrunner
with MPICH may be because it has dual processors.  On CANDYCANE, I have 
experimented with two alternative protocols.  A recent protocol 
called Virtual Interface Architecture (VIA) is being promoted by
Compaq, Intel and Microsoft as a communication architecture for
clusters superior to TCP/IP.  
The National Energy Research Scientific Computing Center (NERSC)
has developed software called M-VIA \cite{MVIA}.
M-VIA supports two types of FastEthernet cards and Packet Engines Gigabit
Ethernet cards.  It will eventually support hardware that is designed to
support the VIA protocol.  For now, a very attractive feature is the
ability to get reduced latency without spending money on more expensive
hardware than FastEthernet.  I have also done a small amount of testing
with the Genoa Active Message Machine (GAMMA) \cite{GAMMA}.  The setup and testing of
this software is not complete, but a ping pong test under GAMMA gave a
latency of 37$\mu$s.  The GAMMA developer estimates that 7 of the 37
$\mu$s come from the switch and that running MPI over GAMMA would add 
an additional 4$\mu$s.

Myrinet hardware is quite popular for high performance Beowulf clusters.
The Roadrunner ``supercluster'' at the Albuquerque High Performance Computer
was used to test Myrinet performance.  When initially tested in August,
1999, the latency was 31-34 $\mu$s.  I recently retested with newer drivers
and found the latency reduced to 16 $\mu$s.

LLNL's Compaq Alpha based cluster is networked with a
Quadrics interconnect.  It is a prototype of the Compaq SC series
supercomputer.  The measured latency there is only 7 $\mu$s.

\section{BENCHMARKS}

The simple performance model presented above can help us predict when the
communication
and floating point are in reasonable balance, but it is no substitute for
real benchmarks.
A web site for MILC benchmarks may be found at
physics.indiana.edu/\~{}sg/milc/benchmark.html.  
Additional graphs and explanations may be found there.
All the benchmarks presented are for single precision Kogut-Susskind
conjugate gradient.  Of course, we run many other codes, but this is one
for which we have results going back several years.

Key variables for the benchmarks are the problem size and number of
CPUs or nodes.  We run benchmarks with $L^4$ sites per CPU and scale the
problem size as the number of CPUs increases.  [In most cases, we double the
dimensions starting with $t$ so that no dimension is more than a factor of
two different from the others.  However, for 4 CPUs we run both 
$L^3 \times 4L$ and $L^2 \times (2L)^2$.]  
We do this because many computers have a
sweet spot for some value of $L$, and in deciding how many nodes to use for
production running, we are usually more interested in running efficiently
than at maximum speed.  The sweet spot comes about because on a single node,
small values of $L$ usually perform best because they take advantage of
cache.  However, as we saw before, small values of $L$ make the most
demands on the network.  As $L$ is increased, the single node performance
decreases, but the degradation from the network may decrease, so
performance may increase until the cache misses become the limiting factor
and performance again decreases.  On PC clusters, the caches tend to be
small and the latency high compared to parallel supercomputers,
and we find that performance usually just continues to increase as $L$ is
increased.  It can be frustrating to run on a small supercomputer
where one is forced to use a larger value of $L$ than optimal.  One gets
decreased performance both because one is running on fewer nodes and
because each node is running more slowly than when running at the sweet
spot.

\begin{figure}[t]
\ifcolor{\epsfxsize=0.99 \hsize\epsffile{fig4.ps}}
\else{\epsfxsize=0.99 \hsize\epsffile{fig4bw.ps}}\fi
\vspace{-28pt}
\caption{Roadrunner benchmarks using a single CPU per node. 
\ifcolor {} \else { (Dashes denote FastEthernet.)} \fi }
\label{fig:roadrunner1}
\end{figure}

\begin{figure}[t]
\ifcolor{\epsfxsize=0.99 \hsize\epsffile{fig5.ps}}
\else{\epsfxsize=0.99 \hsize\epsffile{fig5bw.ps}}\fi
\vspace{-28pt}
\caption{Roadrunner benchmarks using dual CPUs. 
\ifcolor {} \else { (Dashes denote FastEthernet.)} \fi }
\label{fig:roadrunner2}
\end{figure}

The Roadrunner cluster, built by Altatech, has been very useful for
our benchmarking efforts because it has dual processor nodes and one can
use either FastEthernet or Myrinet for the message passing.
In Fig.~\ref{fig:roadrunner1}, we compare the performance using
either network but using only one CPU per node.  
In Fig.~\ref{fig:roadrunner2},
we use both processors.  From this exercise, we were able to determine that
the limitations of the FastEthernet were so great that the second processor
did not improve the cost-effectiveness of the system, but for Myrinet the
second processor did.  (Costs were determined by designing bare bones nodes
with the desired characteristics, not by Altatech's pricing.)

\begin{figure}[t]
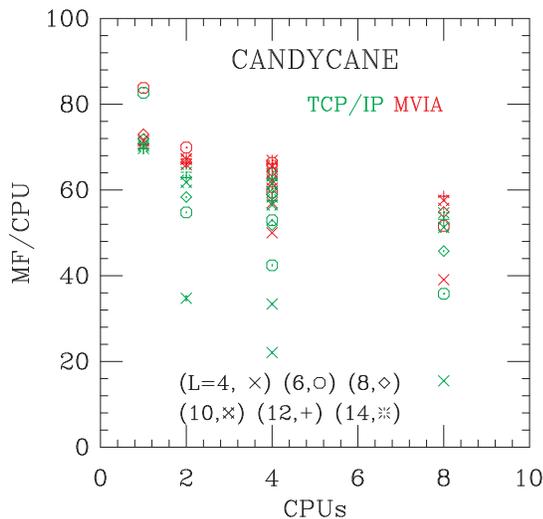

\ifcolor{\epsfxsize=0.99 \hsize\epsffile{fig6.ps}}
\else{\epsfxsize=0.99 \hsize\epsffile{fig6bw.ps}}\fi
\vspace{-28pt}
\caption{Comparison of MVICH (MVIA) and MPICH (TCP/IP) 
using FastEthernet hardware.}
\label{fig:mvich}
\end{figure}

We compare MPICH and MVICH (MPICH running over M-VIA) using the same
FastEthernet hardware in Fig.~\ref{fig:mvich}.  The reduced latency really improves
performance for the smaller values of $L$.  Unfortunately, we don't have
results for more than eight nodes because of a bug in the driver for the
FastEthernet card we have in greatest abundance.  This bug is supposed to
have been fixed, but we have not found the time to install the new drivers
and run the tests.

Moving on to supercomputers, we have results for Cray T3E900, IBM SP with
various speed nodes, SGI Origin 2000, and Sun E10000.  
The Cray T3E does not have a sweet spot.  Its performance
continues to increase as $L$ is increased to 14.  With some assembly code
we get about 75--90 MF per CPU.  For the IBM SP with 4 way SMP Winterhawk
II (375 MHz Power 3) nodes, we find a distinct sweet spot at $L=8$, where
the performance is about 175 MF per CPU.  By $L=12$, however, performance has
dropped to 50 MF per CPU.  These tests were done with up to 64 CPUs.  There
may not be enough bandwidth to memory to support all four CPUs on a node.
We ought to try prefetching data to cache on this computer to see if that
would help.  In a poster by Sonali Tamhankar \cite{SONA}, 
you can find out about our
efforts to speed up the code using OpenMP on the SP and the E10000.
The E10000 also has a sweet spot with peak performance of about 120 MF per
CPU for $L=8$ and 64 CPUs.  By $L=12$, as on the SP, one is down to a
performance level comparable to that of a FastEthernet cluster.

My favorite computer at this point is certainly the LLNL Teracluster,
that has 4-way SMP Alpha EV67 667 MHz nodes and a Quadrics QsNet network.  
It is a cluster, with moderate SMP parallelism and a switched
network; however, the QsNet hardware is not a commodity item at $>$ \$3,000
per node.
We get about 280 MF per CPU at the sweet spot $L=8$.  For $L=12$, performance
is still close to 200 MF per CPU.  These benchmarks use some very old Alpha
assembler code that is far from optimized for the current architecture, so
with some effort, we might be able to do quite a bit better.  (Faster
results have been reported by others on Wilson or Clover inverters.)  
The new NSF Terascale computer to be built at the Pittsburgh Supercomputer
Center will be quite similar to this, but it will have a more advanced
processor.

\begin{figure}[t]
\epsfxsize=0.99 \hsize
\epsffile{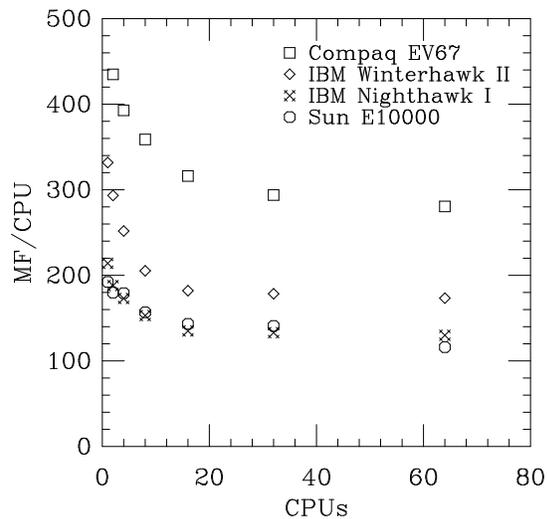}
\vspace{-28pt}
\caption{Comparison of several supercomputers for $8^4$ sites per node.}
\label{fig:supercomputers}
\end{figure}

Figure \ref{fig:supercomputers} summarizes results on 
various supercomputers for $L=8$ on up to
64 CPUs.  For full details visit the MILC benchmark web page.

\section{PRICE-PERFORMANCE RATIOS}

Before presenting any numbers for price-performance ratios, it is essential
to point out some caveats.  It is not trivial to get prices for
supercomputers.  You actually have to take up a salesperson's time to
generate a quote, and that is not a very nice thing to do if your main goal
is to tell the person that the computer is overpriced.  Also, discounts can
vary and details of configuration are not presented here.  On the cluster side,
the configurations were priced based on the lowest prices at www.pricewatch.com
on October 31, 2000.  Each node is configured with 64 MB of ECC memory per
CPU and a 4.3 GB hard drive.
To summarize, the numbers for clusters are up to date, and the numbers
for supercomputers are based on older quotes.  The cluster numbers presented
at the conference were based on node prices over a year old.

We consider cluster nodes with either 450 MHz PII chips as in
Roadrunner, or 733 MHz PIII chips as in Los Lobos.  For the Los Lobos level
system, we are assuming a ServerWorks dual-CPU capable motherboard will have
comparable performance to the IBM Netfinity nodes.  If one of the VIA 
chip based motherboards supports PC133 memory well, it would be a more
cost effective alternative for the single CPU system.
Switch prices are based
on quotes received in January, 2000 or recent advertisements.
Performance expectations are based on Roadrunner or Los Lobos.
(An AMD Athlon based system was considered in Sec.~1.  Performance claims
there were based on RR, although they should be between RR and LL, so
better performance than stated there is very likely.)

To build a single CPU node like Roadrunner, but with 64 MB of memory would
cost \$325.  A dual CPU node, with 128 MB would cost \$527.  If more memory
is desired, it should cost less than \$1 per Megabyte.  Los Lobos style nodes
are \$634 and \$878, for single and dual cpu, respectively.  Per port
FastEthernet switch costs are
about \$56, \$185 and \$240, for 32, 72 and 144 ports, respectively.
Myrinet cost is \$1527 per port and scales linearly up to 128 ports with
the Clos switches.  These prices are for LANai 9, Myrinet 2000 cards.
However, Los Lobos does not yet have the latest switch, so it is not 
running the Myrinet 2000 protocol, which has a peak bandwidth
almost twice as fast as the current value.

For a single CPU RR level system, the price-performance ratio
in \$/MF is
7.2--9.3, 10--13, $\approx$11--15 and 28--31, for FastEthernet with
32, 64, 128 nodes and Myrinet, respectively.  With dual CPUs, the
numbers are 8--11, 12--16, 14--19 and 20--23.  We see that the second
CPU makes the Myrinet based system considerably more cost effective;
however, for FastEthernet, although the marginal cost is small, the
performance gain is not that great either, and the system is less cost
effective.  With the more expensive and higher performance LL level
nodes, FastEthernet cost in \$/MF is about 11, 14 and 17, and
Myrinet is
20--22  for single CPU systems.  With a second processor the Myrinet
number drops to 16--21.  Dual CPU benchmarks have not been run with 
FastEthernet, but the network performance
should be even more of an issue here, and
we expect the cost effectiveness to be somewhat less.

\begin{table}[t]
\caption{Price-performance ratios}
\label{tab:priceperformance}
\begin{tabular}{lc}
\hline
Computer (date of quote)& \$/MF \\
\hline
RR level 1 CPU FE (10/00) & 7--15 \\
RR level 1 CPU Myrinet (10/00) & 28--31 \\
RR level 2 CPU FE (10/00) & 8--19 \\
RR level 2 CPU Myrinet (10/00) & 20--23 \\
LL level 1 CPU FE (10/00) & 11--17 \\
LL level 1 CPU Myrinet (10/00) & 20--22 \\
LL level 2 CPU Myrinet (10/00) & 16--21 \\
\hline
64-node SGI Origin 250 MHz (2/99) & 193\\
44 node Cray T3E (2/99)  & 480 \\
256 node IBM Power 3 SP (2/00) &  166 \\
 \quad with estimated discount &  91 \\
64 CPU Compaq Alpha Server SC & 150 \\
\hline
\end{tabular}
\end{table}

This work was supported by the U.S. DOE under grant DE-FG02-91ER 40661.
Special thanks to the MILC
collaboration, and especially R.~Sugar for reading the manuscript.
We thank the Albuquerque High Performance
Computer Center, Indiana University, LLNL,
National Center for Supercomputing Applications, Pittsburgh
Supercomputer Center and San Diego Supercomputer Center.


\begin{thebibliography}{9}
\bibitem{APS}
The presentations may be found at
physics.indiana.edu/\~{}sg/pccluster.html
\end{thebibliography}
\end{document}